# Current-Perpendicular-to-Plane Giant Magnetoresistance Effect in van der Waals Heterostructures

Xinlu Li[1], Yurong Su[2], Meng Zhu[1], Fanxing Zheng[1], Peina Zhang[1], Jia Zhang[1*], and

Jing-Tao Lü[1*]

[1]*School of Physics and Wuhan National High Magnetic Field Center, Huazhong University of Science and Technology, 430074 Wuhan, China*

[2]*Department of Physics, Hubei University, Wuhan 430062, China*

* jiazhang@hust.edu.cn;

*jtlu@hust.edu.cn;

**Abstract:**

Spin-dependent transport in a full van der Waals (vdW) giant magnetoresistance (GMR) junctions with the structure of $Fe_3GeTe_2$/$X$$Te_2$/$Fe_3GeTe_2$ ($X$ = Pt, Pd) has been investigated by using first-principles calculations. The ballistic conductance, magnetoresistance (MR) and resistance-area product (RA) have been calculated in a current-perpendicular-to-plane (CPP) geometry. A giant magnetoresistance of around 2000% and RA less than 0.3 Ω $\mu m^2$ have been found in the proposed vdW CPP GMR. In addition, the spin-orbit coupling effect on transport and anisotropy magnetoresistance (AMR) has also been investigated. The calculated AMR is found to be around 20% in $Fe_3GeTe_2$/trilayer-$PdTe_2$/$Fe_3GeTe_2$ CPP GMR. Both GMR and AMR in the proposed vdW CPP GMR mainly originate from the bulk electronic structure properties of $Fe_3GeTe_2$. This work demonstrates a vdW CPP GMR with superior advantages including perpendicular magnetic anisotropy, large GMR, low RA as well as sizable AMR may stimulate future experimental explorations and should be appealing for their applications in spintronic devices including magnetic sensor and memory.


# I. INTRODUCTION

Giant magnetoresistance (GMR) effect in magnetic metallic multilayers was discovered by A. Fert[1] and P. Grünberg[2] in the 1980s. Since then, metallic GMR structure, for instance, $[Co/Cu]_N$ multilayers with magnetoresistance ratio (MR) over tens of percent at room temperature, has been widely studied and applied in modern spintronic devices. For example, GMR has been used as a read head in a hard disk drive (HDD)[3], and a storage cell in a magnetic random-access memory (MRAM)[4]. By replacing the nonmagnetic metallic spacer with a nanometer thick insulating barrier *e.g.* amorphous $AlO_x$[5], and later single-crystalline MgO barrier, the so-called magnetic tunnel junctions (MTJs) have been shown to have larger tunneling magnetoresistance (TMR) exceeding several hundreds of percent at room temperature[6][7][8][9][10].

Nevertheless, for the applications in next-generation higher density HDD ( > 2 Tbits/in$^2$) and MRAM( > 10 Gbits/in$^2$), there are significant challenges for conventional GMR and TMR junctions to satisfy the following criteria simultaneously[11][12]: (i) Larger MR for high readout signal. (ii) Low resistance-area product (RA) for impedance matching and power consumption reduction. (iii) Perpendicular magnetic anisotropy of magnetic layers to achieve faster magnetization switching and minimize stray field between neighboring cells (and therefore high area density) in MRAM. For example, for MRAM with a density of around 10 Gbits/in$^2$, it requires a storage cell that has RA < 3.5 Ω $\mu m^2$ and MR > 100%[5] and for HDD with an area density of 2 Tbits/in$^2$, it requires similar MR but an order of magnitude smaller RA. Unfortunately, at present, neither GMR junctions nor MTJs can meet the above requirements. The metallic GMR junctions have low RA, but MR is also low (typically less than 100%). For TMR junctions, the MR is usually high enough, but the RA is inevitably large due to the electron tunneling process across the insulating barrier. Those restrictions originate from the intrinsic electronic properties of conventional materials.

Stimulated by recently discovered vdW materials, one may be able to build magnetic vdW heterostructures with appealing properties. Indeed, heterostructures based on magnetic vdW materials have been revealed to have intersting functionalities. Recently, it has been reported that by using $CrI_3$ as spin filter barrier, the Graphite/$CrI_3$/Graphite vdW tunnel junctions exhibit a record high magnetoresistance over thousands of percent at low temperature[13][14][15]. What's more, MTJs using $Fe_3GeTe_2$ as ferromagnetic electrodes have been investigated both experimentally[16] and theoretically[17][18]. Besides, large

TMR induced by spin-filter effect has been reported in transitional metal dichalcogenides MTJ[19].

Here, we build a full vdW GMR junction and consider the CPP (current-perpendicular-to-plane) transport geometry with structure of $Fe_3GeTe_2$/$X$$Te_2$/$Fe_3GeTe_2$ ($X$ = Pt, Pd) by using a class of type-II Dirac semimetal $X$$Te_2$($X$ = Pt, Pd)[20][21][22][23][24] as the nonmagnetic metallic spacer. $Fe_3GeTe_2$ is chosen as magnetic electrode since among the known 2D ferromagnetic vdW metals, $Fe_3GeTe_2$ exhibits relatively high Curie temperature of around 220 K[25][26][27][28][29] and can be further enhanced above room temperature by ionic gating[30]. In addition, since both $PtTe_2$ and $PdTe_2$ include heavy elements, the spin-orbit coupling may be sufficient to produce appreciable anisotropic transport phenomena. We investigate the ballistic transport phenomena, including giant magnetoresistance and anisotropic magnetoresistance effect in the proposed vdW CPP GMR by first-principles calculations.

## II. CALCULATION METHODS

The first-principles calculations are performed at the density functional theory level by using the generalized gradient approximation (GGA)[31] of exchange correlation potential and the ultrasoft pseudopotential[32] as implemented in Quantum ESPRESSO[33]. A Monkhorst-Pack **k**-point mesh of 12 × 12 × 1 is used and the cutoff energies for wave function and charge density are set to 40 Ry and 320 Ry for self-consistence calculations of the $Fe_3GeTe_2$/$X$$Te_2$/$Fe_3GeTe_2$ ($X$ = Pt, Pd) supercells. The vdW interaction has been taken into account by employing the DFT-D3 scheme[34][35]. Before conducting the electron transport calculation of CPP GMR, we calculated the magnetic crystalline anisotropy (MCA) energy of bulk $Fe_3GeTe_2$ by using force theorem with employing Vienna Ab initio Simulation Package (VASP)[36]. The MCA in force theorem is defined as the band energy difference between various magnetization orientations. In order to achieve converged MCA, a much denser **k**-point mesh of 20 × 20 × 5 has been adopted. The calculated MCA (defined as the energy difference between magnetization along $c$ and $a$-axis) of bulk $Fe_3GeTe_2$ is around 1.03 meV/Fe, which agrees well with previous theoretical and experimental results[37][38], confirming that the vdW CPP GMR with $Fe_3GeTe_2$ electrode should have perpendicular magnetization easy axis.

The interface atomic structure of vdW CPP GMR is established by matching the ( 1 × 1 ) in-plane cell of $X$$Te_2$ ($X$ = Pt, Pd) and $Fe_3GeTe_2$ electrodes due to the small lattice

mismatch[39]. The in-plane lattice constant of CPP GMR is fixed to the experimental in-plane lattice constant of Fe$_3$GeTe$_2$ ($a = b = 3.991$ Å). In heterostructures, the stacking configuration and interlayer distance between Fe$_3$GeTe$_2$ and PtTe$_2$ (PdTe$_2$) are adjusted to minimize the total energy. It is found that the interface atomic configurations are most energy favorable when the Pt (Pd) and Te atoms in the interfacial PtTe$_2$ (PdTe$_2$) layer sit at the same in-plane sites of Fe and Ge atoms in interfacial Fe$_3$GeTe$_2$ layer respectively. (Please see Supplementary Note 2 for detailed total energy calculations of various interface stacking configurations[39]). The Fe$_3$GeTe$_2$/trilayer $X$Te$_2$/Fe$_3$GeTe$_2$ ($X$ = Pt, Pd) CPP GMR structure is shown in Fig. 1(a) and the atomic positions in the junctions have been fully relaxed until the forces on each atom are less than $10^{-4}$ Ry/a$_B$ ≈ 2.57 meV/Å, where 1 Ry = 13.6 eV and a$_B$ = 0.529 Å is the Bohr radius. The equilibrium interlayer distance between Fe$_3$GeTe$_2$ and PtTe$_2$ (PdTe$_2$) are found to be 2.747 Å (2.682 Å), respectively, which is typical vdW type of interface spacing.

After building the Fe$_3$GeTe$_2$/$X$Te$_2$/Fe$_3$GeTe$_2$ ($X$ = Pt, Pd) junctions as scattering region, the self-consistent calculations of the Fe$_3$GeTe$_2$ electrode and scattering region have been conducted separately. Then the $\mathbf{k}_\| = (k_x, k_y)$ resolved electron transmission is calculated by matching the wave functions of Fe$_3$GeTe$_2$ electrodes and scattering region with 200 × 200 $\mathbf{k}_\|$ mesh in two-dimensional Brillouin zone (2D BZ). The ballistic conductance of the CPP GMR can be obtained by summarizing the transmission over 2D BZ based on the Landauer formula[40][41][42] as follows:

$$G = \frac{e^2}{h} \sum_{\mathbf{k}_\|} T(\mathbf{k}_\|) \qquad (1)$$

where $T(\mathbf{k}_\|)$ is the electron transmission probability at Fermi energy, $\mathbf{k}_\|$ is the wave vector, $e$ is the elementary charge and $h$ is the Planck constant. When the GMR is calculated in the absence of spin-orbit coupling (SOC), the conductance can be written in spin-resolved form as[9][10]:

$$G_\sigma = \frac{e^2}{h} \sum_{\mathbf{k}_\|} T_\sigma(\mathbf{k}_\|) \qquad (2)$$

where $\sigma$ is the spin-index.

When the anisotropic magnetoresistance (AMR) is studied in the presence of SOC, the Bloch states are no longer eigenstates of spin operator, and each Bloch state has mixed spin

up and spin down components. Therefore, in this situation the conductance for each magnetic configuration is calculated by Equation (1). The RA of CPP GMR can be evaluated from the calculated transmission by its definition[39].

## III. RESULTS AND DISCUSSIONS

The crystal structures of bulk $X$Te$_2$ ($X$ = Pt/Pd) and Fe$_3$GeTe$_2$ are shown in Fig. 1(a) and (b), respectively. Fig. 1(c) depicts their first BZ. The band structures of bulk Fe$_3$GeTe$_2$, PtTe$_2$ and PdTe$_2$ along the $\Gamma$-$A$ direction (transport direction) are shown in Fig. 1(d-g). For Fe$_3$GeTe$_2$, the spin up band structure crossing Fermi level belongs to $\Delta_1$ symmetry, which is mainly composed of Fe-$d_z^2$ orbital (Please see Fig. S2 in Supplementary Note 4 for atomic weight projection of band structures[39]). The spin down band structure also has similar $\Delta_1$ symmetry around Fermi energy, which consists of Fe-$d_z^2$ and Ge-$p_z$ orbitals. On the other hand, for PdTe$_2$ and PtTe$_2$, the energy bands at Fermi energy also have $\Delta_1$ symmetry, which are mainly contributed by Te-$p_z$ orbital. In CPP transport geometry, the Bloch states with $\Delta_1$ symmetry at Fermi energy should have relatively large transmission due to its itinerant feature of the atomic orbital compositions[9]. This agrees with the calculated results where the transmissions around $\bar{\Gamma}$ point are generally very large for both spin channels for Fe$_3$GeTe$_2$/trilayer $X$Te$_2$/Fe$_3$GeTe$_2$ ($X$=Pt, Pd) CPP GMR.

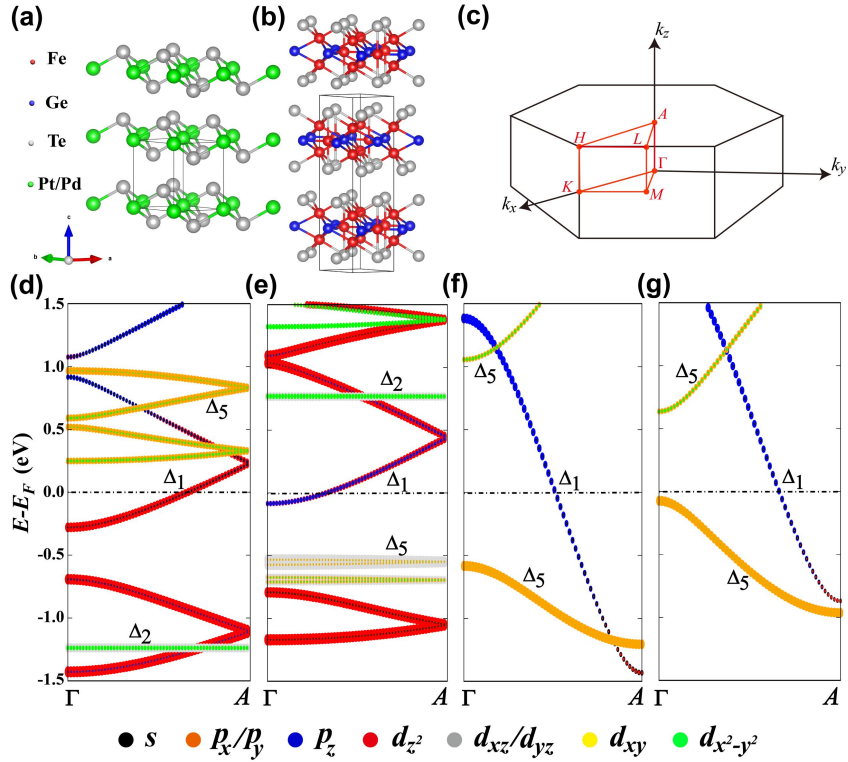

Fig. 1 (a), (b) Crystal structures of bulk $X$Te$_2$ ($X$ = Pt/Pd) and Fe$_3$GeTe$_2$. Black boxes in (a), (b) indicate the bulk unit cell. (c) Brillouin zone and high-symmetry **k**-path of a hexagonal lattice. (d), (e) Spin-up and spin-down band structures with orbital projections for Fe$_3$GeTe$_2$ along Γ-$A$ direction. (f), (g) Corresponding projected band structures for bulk PtTe$_2$ and PdTe$_2$. Size of symbols in (d)-(g) is proportional to the projection weight of each orbital.

The 2D Fermi surfaces (FS) of electrodes, which indicate the distribution of available travelling Bloch states in 2D BZ, play a decisive role in electron transport in the CPP GMR. Therefore, we calculated the 2D FS of bulk Fe$_3$GeTe$_2$ by calculating the transmission of identical Fe$_3$GeTe$_2$/Fe$_3$GeTe$_2$/Fe$_3$GeTe$_2$ sandwiched structures. The calculated spin polarized 2D FS of Fe$_3$GeTe$_2$ are shown in Fig. 2(b) and (c). By comparing Fig. 2(b) and Fig. 2(c), one can notice that the 2D FS for two spin channels are remarkably different. For spin up channels, the Bloch states are available in most regions of the 2D BZ, while for spin down channels, only a small area around $\bar{\Gamma}$ ($k_x$ = 0, $k_y$ = 0) and $\bar{K}$ points have available Bloch states. Such largely mismatched and highly spin-polarized 2D FS between two spin channels would greatly reduce the transmission for antiparallel magnetization configuration, which might instead ensure high GMR ratio in the proposed vdW CPP GMR.

The illustration of atomic structure for Fe$_3$GeTe$_2$/trilayer $X$Te$_2$/Fe$_3$GeTe$_2$ ($X$ = Pt, Pd) CPP GMR is shown in Fig. 2(a). We then explicitly calculated the electron transmission in the CPP GMR junctions for parallel (P) and antiparallel (AP) magnetization configurations of two Fe$_3$GeTe$_2$ electrodes. As seen from Fig. 2(d-i), the electron transmissions for spin up and spin down channels generally resemble the shape of corresponding 2D FS of Fe$_3$GeTe$_2$ (Fig. 2(b) and 2(c)) but with modulated transmission at each **k**$_\parallel$ point. For instance, the transmission of spin up electron for P configuration is consistent with the spin up 2D FS (Fig. 2(b)), and the transmission of spin down electron for P configuration as well as the transmission for AP configuration (Fig. 2(f) and Fig. 2(i)) agree with the 2D FS of spin down channels shown in Fig. 2(c). In a large area of 2D BZ, the transmissions of spin up channels for P configuration are larger than 0.1, which results in a much larger total transmission (conductance) than AP configurations. In addition, the band symmetry of electron in electrodes at each **k**$_\parallel$ point is also important for determining the transmission through the CPP GMR.

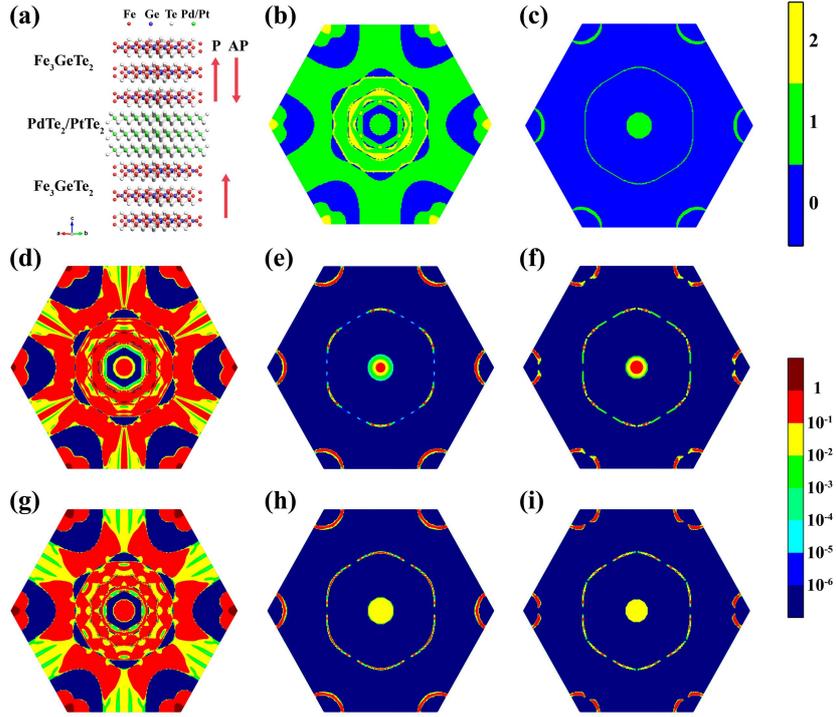

Fig. 2(a) Atomic structure of Fe$_3$GeTe$_2$/trilayer $X$Te$_2$/Fe$_3$GeTe$_2$ ($X$ = Pt, Pd) CPP GMR. Red arrows indicate the relative magnetization orientations of two Fe$_3$GeTe$_2$ electrodes. (b),(c) 2D FS of spin up and spin down channels for bulk Fe$_3$GeTe$_2$. The color scale alongside (c) indicates the number of available Bloch states for two spin channels at each $\mathbf{k}_\parallel = (k_x, k_y)$ point. (d),(e) Spin-up and spin-down transmissions for Fe$_3$GeTe$_2$/trilayer PdTe$_2$/Fe$_3$GeTe$_2$ CPP GMR for P configuration. (f) Transmission for AP configuration for spin up (or spin down) channels. Figures from (g)-(i) Same but for Fe$_3$GeTe$_2$/trilayer PtTe$_2$/Fe$_3$GeTe$_2$ CPP GMR. Color scale alongside (f),(i) indicates the magnitude of transmissions shown for (d)-(i) on a logarithmic scale.

The calculated electron transmissions, RA and the corresponding GMR ratio (defined as $[(G_P/G_{AP})-1]\times 100\%$, where $G_P$ and $G_{AP}$ are the conductance for P and AP magnetic configurations) for Fe$_3$GeTe$_2$/trilayer $X$Te$_2$/Fe$_3$GeTe$_2$ ($X$ = Pt, Pd) CPP GMR are listed in Table 1. We have ignored the SOC effect when the giant magnetoresistance (GMR) is investigated (see Supplementary Note 6 for the discussions on SOC effect on GMR[39]), since GMR ratio mainly depends on the relative magnetization orientations and the spin-polarized band structures of the two magnetic electrodes. Generally, GMR ratio of around 2000% and RA of less than 0.3 Ω $\mu$m$^2$ have been obtained in the proposed vdW CPP GMR. In comparison to the previously studied Fe$_3$GeTe$_2$/Graphene/Fe$_3$GeTe$_2$ tunneling magnetoresistance structure where RA is around 5 Ω $\mu$m$^2$[17], it suggests that the proposed metallic CPP GMR with low RA and high magnetoresistance may be more promising for device application. It's worth noting that the transport properties in the CPP GMR do not

change significantly either by using trilayer-PtTe$_2$ or trilayer PdTe$_2$ as metallic spacer layers. This is not surprising, since the interface interaction between Fe$_3$GeTe$_2$ electrode and spacer layers are weak vdW type, which makes the transport properties of CPP GMR mostly determined by Fe$_3$GeTe$_2$ electrode rather than spacer layers and the interfaces. And in consequence, the large GMR ratio in this class of vdW CPP GMR could be mainly attributed to the highly spin-polarized electronic structures of Fe$_3$GeTe$_2$. That is the significantly distinct 2D FS for two spins as we discussed.

The origin of large GMR ratio in the studied vdW CPP GMR differs from the conventional GMR multilayer structures. For GMR multilayer structures, both the bulk and interface spin-dependent scattering are crucial[43]. Typically, the experimentally reported GMR ratio of conventional [Co/Cu]$_N$ and [Fe/Cr]$_N$ multilayers is around tens of percent at room temperature[44][45]. Besides, in conventional CPP GMR using high spin-polarized Heusler alloys as magnetic electrodes, the GMR ratio can be further enhanced[46][47][48]. In Table 2 we list the GMR ratio and RA for typical conventional CPP GMR structures. The largest GMR ratio obtained in CPP GMR with Heusler alloy electrodes is found to be around 73% at room temperature which is attributed to the enhanced interfacial spin asymmetry scattering[48].

Recently, 2D materials have also been adopted in CPP GMR. For example, in the Fe/MoS$_2$/Fe CPP GMR by using transition metal dichalcogenide (TMDC) semiconductor MoS$_2$ as a spacer layer, the first-principles calculations predict a GMR ratio up to 300%[49]. Experimentally, GMR ratio is reported to be less than 1% at low temperature (T = 10 K) and with an extremely large RA of around 100 Ω $\mu$m$^2$ in NiFe/MoS$_2$/NiFe CPP GMR[50]. More recently, vdW CPP GMR with structure of Fe$_3$GeTe$_2$/MoS$_2$/Fe$_3$GeTe$_2$ has also been experimentally fabricated and the GMR ratio is found to be around 3.1% at 10 K[51]. The achieved low experimental GMR ratio in such CPP GMR by using MoS$_2$ as spacer layer may be attributed to several reasons. First, although the contact between magnetic electrodes and MoS$_2$ spacer is metallic due to the strong interface hybridization, the resultant RA should be large since much fewer energy bands are present at the Fermi level compared to metallic spacers like PtTe$_2$ and PdTe$_2$. Second, semiconductor may not be a good spacer layer candidate for GMR, since metal-semiconductor contact may not be ideal and could produce Fermi energy pinning effect[52] which are detrimental for spin dependent transport.

Experimentally, the crystal structure in vdW heterostructures may not be perfect because of the weak interlayer interaction. For example, there may be interlayer

misalignments and relative rotations between Fe$_3$GeTe$_2$ electrode and PtTe$_2$ (PdTe$_2$) spacer layer. Such structural imperfections may impact the transport properties in the studied vdW CPP GMR. As our previous theoretical model[17] shows that the magnetoresistance shows weak dependence on the interlayer translation, while an order of magnitude decrease of magnetoresistance is found when the relative rotation of the two Fe$_3$GeTe$_2$ electrodes occurs. At finite temperature, the GMR ratio may decrease with the increase of temperature because of the demagnetization of magnetic vdW electrodes and possible spin-flipped scattering by magnons. As we discussed before, the electronic structures of the magnetic electrode will mainly determine the GMR ratio. Therefore, the key for achieving high experimental GMR value in vdW magnetic heterostructures at finite temperature is to explore vdW ferromagnetic materials with Curie temperature well above room temperature and high spin-polarization (ideally half-metallic ferromagnets with only one type of conducting spin channels at Fermi energy).

Table 1. The calculated spin-dependent electron transmission $T_\uparrow$ and $T_\downarrow$, RA for P and AP magnetization alignments of two Fe$_3$GeTe$_2$ layers and the GMR ratio in Fe$_3$GeTe$_2$/trilayer $X$Te$_2$/Fe$_3$GeTe$_2$ ($X$ = Pt, Pd) vdW CPP GMR.

| | $M_{\uparrow\uparrow}$ (Parallel Magnetization) | | | | $M_{\uparrow\downarrow}$ (Antiparallel Magnetization) | | | |
|---|---|---|---|---|---|---|---|---|
| | Spin up $T_\uparrow$ | Spin down $T_\downarrow$ | $T$ (=$T_\uparrow$+$T_\downarrow$) | RA (mΩ μm$^2$) | Spin up $T_\uparrow$/Spin down $T_\downarrow$ | $T$ (=$T_\uparrow$+$T_\downarrow$) | RA (mΩ μm$^2$) | GMR ratio |
| trilayer-PtTe$_2$ | 3.25×10$^{-1}$ | 8.99×10$^{-3}$ | 3.34×10$^{-1}$ | 10.7 | 8.04×10$^{-3}$ | 1.60×10$^{-2}$ | 222.5 | 1979% |
| trilayer-PdTe$_2$ | 3.25×10$^{-1}$ | 9.07×10$^{-3}$ | 3.34×10$^{-1}$ | 10.7 | 8.28×10$^{-3}$ | 1.66×10$^{-2}$ | 214.5 | 1905% |

Table 2. Experimental RA, ΔRA (defined as (RA)$_{AP}$-(RA)$_P$) and GMR ratio for typical CPP GMR.

| Structures of CPP GMR | RA (mΩ μm$^2$) | ΔRA (mΩ μm$^2$) | GMR ratio | References |
|---|---|---|---|---|
| [Co/Cu]$_{180}$; [Fe/Cr]$_{100}$ | -- | -- | 56%; 14% (T=300 K) | [45] |
| Co$_2$Fe$_{0.4}$Mn$_{0.6}$Si/Ag/ Co$_2$Fe$_{0.4}$Mn$_{0.6}$Si | 17.2~369.2 (AP state) | -- | 62~75% (T=300 K) | [47] |
| Co$_2$Fe$_{0.4}$Mn$_{0.6}$Si/Ag-Mg/ Co$_2$Fe$_{0.4}$Mn$_{0.6}$Si | 37 (P state) | 27 | 73% (T=300 K) | [48] |
| NiFe/MoS$_2$/NiFe | ~10$^5$ | -- | 0.4% (T=10 K) | [50] |
| Fe$_3$GeTe$_2$/MoS$_2$/Fe$_3$GeTe$_2$ | -- | 841.9 | 3.1% | [51] |



The spin-orbit coupling (SOC) effect plays an essential role in the appearing topological band structures of $Fe_3GeTe_2$ and $XTe_2$($X$ = Pd, Pt) and the triggering magnetic anisotropy of $Fe_3GeTe_2$. For example, when SOC has been taken into account, $Fe_3GeTe_2$ is found to be nodal line semimetal[28] and $XTe_2$($X$ = Pd, Pt) are type-II Dirac semimetals[20][21][22]. Spin-orbit coupling may also have a notable impact on electron transport in the proposed vdW CPP GMR and produce anisotropic conductance dependent on the magnetization orientations of $Fe_3GeTe_2$.

The band structures of $Fe_3GeTe_2$ along Γ-$A$ high symmetry line (parallel to the transport direction) for $M_z$ and $M_x$ cases are shown in Fig. 3(a) and (b). It is clear that the band dispersion is strongly coupled to the magnetization orientation due to the sizable SOC effect in $Fe_3GeTe_2$. Similar magnetization direction dependence of band structures in $Fe_3GeTe_2$ along other high symmetry lines, for instance, $M$-$K$, $K$-Γ and Γ-$H$ in Brillouin Zone has also been theoretically discussed[28]. Consequently, the Bloch states of $Fe_3GeTe_2$ at each $\mathbf{k}_\parallel$ point in 2D BZ possess different band features like crystal momentum $k_z$ and electron velocity $v_z$ along transport direction for $M_z$ and $M_x$ cases, which leads to anisotropic transmission probability in CPP GMR. We have also investigated the anisotropic magnetoresistance (AMR) effect of $Fe_3GeTe_2$/trilayer $PdTe_2$/$Fe_3GeTe_2$ CPP GMR by calculating the electron transmission in the presence of SOC when the magnetization directions of two $Fe_3GeTe_2$ electrodes are fixed to be parallel but rotate simultaneously along $z$ and $x$ directions. The details for transport calculations in the presence of SOC can be found in Supplementary Note 7[39]. The transmission of CPP GMR at $\bar{\Gamma}$ point as a function of energy is calculated and shown in Fig. 3(c). It is clear that in a wide energy window, the transmission for $M_z$ and $M_x$ are remarkably different. Especially, for energy slightly above Fermi energy, a stronger anisotropic transmission is present which indicates a possible larger AMR at that energy window.

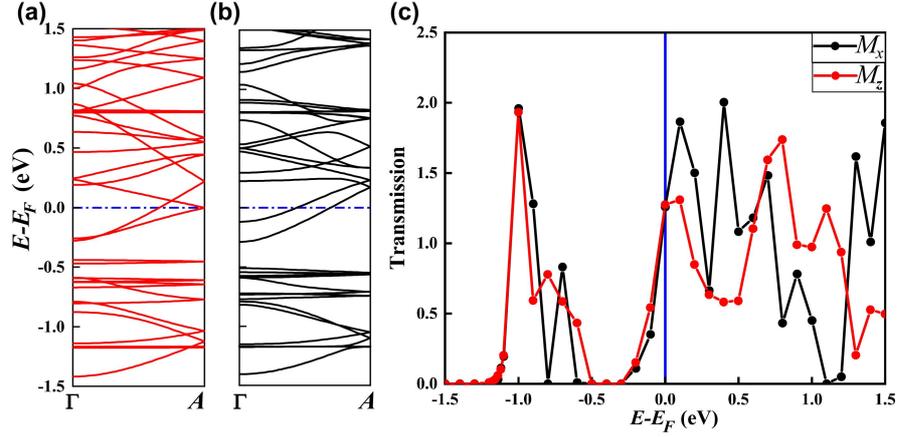

Fig. 3 (a), (b) Energy-band structures of bulk $Fe_3GeTe_2$ along $\Gamma$-$A$ direction for $M_z$ (red lines) and $M_x$ (black lines) magnetization states. (c) Transmission of $Fe_3GeTe_2$/trilayer $PdTe_2$/$Fe_3GeTe_2$ CPP GMR as a function of energy at $\mathbf{k}_{\parallel} = (0,0)$ for $M_z$ and $M_x$ cases. Fermi energy is set to zero. Notably, over some energy ranges, transmission is larger than one, since more than one Bloch state is present at the corresponding energy.

Similar, prior to calculating the transmission, we calculated and plotted the three-dimensional Fermi surface (3D) FS of bulk $Fe_3GeTe_2$ for magnetization along $z$ ($M_z$) and $x$ ($M_x$) axis. The 3D FS of $Fe_3GeTe_2$ are shown in Fig. 4(b) and (c), which exhibit different **k**-symmetry in BZ. For the $M_z$ case, the magnetic point group of $Fe_3GeTe_2$ is $D_{6h}$ (6/mmm), which has 24 symmetry operations in total, while for $M_x$ case, the magnetic point group belongs to $D_{2h}$(mmm) with lower symmetry. The irreducible 2D BZ where the transmission has been explicitly calculated for $M_z$ and $M_x$ cases are indicated by the red frames shown in Fig. 4(d) and (e). Most importantly, the 2D FS for $M_z$ case (Fig. 4(d)) covers less area than the $M_x$ case (Fig. 4(e)) around $\overline{\Gamma}$ point. As we have discussed previously, the Bloch states with $\Delta_1$ band symmetry around $\overline{\Gamma}$ points should have relatively large transmission as it is shown in Fig. 4(f) and (g). In consequence, the evaluation of the FS around $\overline{\Gamma}$ point by rotating magnetization of $Fe_3GeTe_2$ leads to an anisotropic conductance and a larger total transmission for $M_x$ than $M_z$ case.

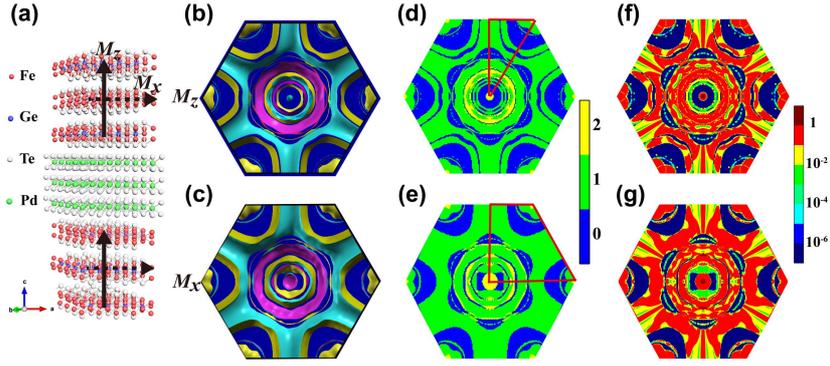

Fig. 4. (a) Atomic structure of Fe$_3$GeTe$_2$/trilayer PdTe$_2$/Fe$_3$GeTe$_2$ CPP GMR. Black arrows indicate the magnetization directions of two Fe$_3$GeTe$_2$ electrodes. Top view of 3D FS (b), (c); 2D FS (d), (e) of bulk Fe$_3$GeTe$_2$ and the electron transmission distribution of the CPP GMR (f), (g) in the presence of SOC for $M_z$ and $M_x$ cases. Colors in (b), (c) indicate Fermi surfaces belonging to different bands. Red frame lines in (d), (e) show the irreducible BZ for calculating 2D FS and transmission. Color scales along the side of (d), (e) and (f), (g) indicate available Bloch states and transmission, respectively.

The calculated transmission, RA and AMR ratio (defined as: $[G(M_x)/G(M_z)-1]\times 100\%$) in Fe$_3$GeTe$_2$/trilayer PdTe$_2$/Fe$_3$GeTe$_2$ CPP GMR are listed in Table 3. The transmission of CPP GMR for $M_x$ case is larger than $M_z$ case, which leads to an AMR of around 20%. The origin of AMR in vdW CPP GMR can also be mainly attributed to the bulk electronic structure evaluation of Fe$_3$GeTe$_2$ for different magnetization orientations as we have discussed previously. In contrast, the origin of AMR is different in conventional AMR structure. For instance, in magnetic tunnel junctions, the TAMR (tunneling anisotropic magnetoresistance) generally originates from the surface and interface resonant states[53][54]. Recently, antisymmetric magnetoresistance effect has been reported in Fe$_3$GeTe$_2$/Graphite/Fe$_3$GeTe$_2$ trilayer heterostructures in a CIP (Current-In-Plane) geometry which originates from a different mechanism. The spin momentum locking induced spin-polarized current at the graphite/Fe$_3$GeTe$_2$ interface is found to be responsible for the presence of MR in this system[55].

Table 3. The calculated electron transmission, RA and AMR in Fe$_3$GeTe$_2$/trilayer PdTe$_2$/Fe$_3$GeTe$_2$ CPP GMR.

| $M_z$ | | $M_x$ | | |
|---|---|---|---|---|
| Transmission | RA (mΩ $\mu$m$^2$) | Transmission | RA (mΩ $\mu$m$^2$) | **AMR** |
| 2.20×10$^{-1}$ | 16.18 | 2.65×10$^{-1}$ | 13.43 | 20.5% |

The full vdW heterostructures proposed in this work should be fabricated more easily and retain better crystal quality with fewer interface defects including interface alloy and surface roughness. in comparison to traditional CPP GMR. In the proposed CPP GMR, the parallel and antiparallel magnetic orientations of two $Fe_3GeTe_2$ electrodes can be realized by employing different coercivities of two $Fe_3GeTe_2$ layers by layer thickness and film shape engineering. An alternative way to realize the P and AP states of CPP GMR is to pin one of the $Fe_3GeTe_2$ layers with an adjacent antiferromagnetic layer[56][57]. In the case of anisotropic magnetoresistance, the different magnetic moment orientations of two $Fe_3GeTe_2$ can be achieved by applying sufficiently large magnetic field along various directions[54].

## IV. SUMMARY

In summary, the spin-dependent transport properties in a full vdW CPP GMR with the structure of $Fe_3GeTe_2$/trilayer $X$$Te_2$/$Fe_3GeTe_2$ ($X$=Pt, Pd) have been studied by employing first-principles calculations. Giant GMR ratio of around 2000%, low RA less than 0.3 Ω $\mu m^2$ and perpendicular magnetization anisotropy have been predicted simultaneously in those vdW metallic CPP GMR. When spin-orbit coupling is taken into account, the $Fe_3GeTe_2$/trilayer $PdTe_2$/$Fe_3GeTe_2$ CPP GMR is found to have an AMR value of around 20%. The mechanism for the GMR effect in the studied vdW CPP GMR mainly relies on the spin-polarized electronic structure of the ferromagnetic $Fe_3GeTe_2$ electrodes, whereas the AMR effect originates from anisotropic electronic structure of $Fe_3GeTe_2$ along different magnetization. Our calculation results demonstrate the proposed vdW CPP GMR having perpendicular magnetic anisotropy, giant GMR ratio, low RA as well as moderate AMR should be promising for device applications including magnetic sensor, memory and so forth. This work may stimulate future experimental explorations on magnetoresistance effect in vdW type of CPP GMR.

## ACKNOWLEDGMENTS

J. Z. and J. T. L. acknowledge the support from the National Natural Science Foundation of China (grants No. 11704135, No. 21873033), the program for HUST academic frontier youth team. The computations in this work are partly performed on the Platform for Data-Driven Computational Materials Discovery at the Songshan Lake Materials Laboratory, Dongguan, China.